\documentclass[aip, rsi, amsmath, amssymb, reprint]{revtex4-1}

\usepackage{CJK}

\usepackage{graphicx}
\usepackage{dcolumn}
\usepackage{bm}
\usepackage{siunitx}
\usepackage{braket}
\usepackage{hyperref}

\begin{document}
\begin {CJK*} {GB} { } 

\preprint{}

\title[]{Sub-picotesla widely-tunable atomic magnetometer operating at room-temperature in unshielded environments}

\author{Cameron Deans}
\author{Luca Marmugi}
\email{l.marmugi@ucl.ac.uk}
\author{Ferruccio Renzoni}
\affiliation{Department of Physics and Astronomy, University College London, Gower Street, London WC1E 6BT, United Kingdom}

\date{\today}

\begin{abstract}
We report on a single-channel rubidium radio-frequency atomic magnetometer operating in unshielded environments and near room temperature with a measured sensitivity of \SI{130}{\femto\tesla\per\sqrt\hertz}. We demonstrate consistent, narrow-bandwidth operation across the \SI{}{\kilo\hertz}~--~\SI{}{\mega\hertz} band, corresponding to three orders of magnitude of magnetic field amplitude. A compensation coil system controlled by a feedback loop actively and automatically stabilizes the magnetic field around the sensor. We measure a reduction of the \SI{50}{\hertz} noise contribution by an order of magnitude. The small effective sensor volume, \SI{57}{\milli\metre\cubed}, increases the spatial resolution of the measurements. Low temperature operation, without any magnetic shielding, coupled with the broad tunability, and low beam power, dramatically extends the range of potential field applications for our device.\\

\vskip 20pt

\begin{center}
This is a preprint version of the article appeared in Review of Scientific Instruments:\\
C. Deans, L. Marmugi, F. Renzoni, Rev. Sci. Instr. \textbf{89}, 08311 (2018) DOI: \href{https://doi.org/10.1063/1.5026769}{10.1063/1.5026769}.
\end{center}
\end{abstract}

                             

\maketitle
\end{CJK*}

\section{\label{sec:Intro}Introduction}

The ultra-sensitive detection of magnetic fields is a requirement for an increasing number of applications and technologies. Atomic magnetometers (AMs) \cite{budker2007optical}, featuring the optical pumping and interrogation of an alkali vapor, compete with superconducting quantum interference devices (SQUIDs) \cite{kleiner2004superconducting, fagaly2006superconducting} for record sensitivities, without the requirement of cryogenic temperatures. This represents an advantage in terms of functionality and flexibility. Further advantages are: low power consumption and running costs, and the potential for miniaturization -- leading to hand-held devices \cite{schwindt2004chip, balabas2006magnetometry, shah2007subpicotesla, griffith2010femtotesla}. 

Radio-frequency atomic magnetometers (RF-AMs) \cite{savukov2005tunable, savukov2014ultra} are tunable over a wide range of operation frequencies and have applications in magnetic resonance imaging (MRI) \cite{xu2006magnetic, savukov2009mri, savukov2013magnetic}, nuclear quadrupole resonance (NQR) \cite{lee2006subfemtotesla, cooper2016atomic}, nuclear magnetic resonance (NMR) \cite{savukov2007detection, bevilacqua2009all, bevilacqua2017simultaneous}, electromagnetic induction imaging (EII) \cite{deans2016electromagnetic, deans2017through, wickenbrock2016eddy}, along with medical applications such as magnetocardiography (MCG) \cite{belfi2007cesium, knappe2015fetal}.

Several parameters can be used to compare the suitability of competing RF-AMs implementations to practical applications. These include sensitivity, tunability of operation frequency, operation bandwidth, and modularity. The sensitivity crucially depends on the magnetic field noise. Therefore, a significant difference in performance is associated with AMs in shielded and unscreened environments.

The majority of previous RF-AMs designs feature multiple layers of mu-metal shielding enclosing the sensor \cite{savukov2005tunable, lee2006subfemtotesla, ledbetter2007detection, chalupczak2012room, savukov2014ultra}. This results in an increased sensitivity, with the atomic vapor protected against magnetic field noise. However, the cost and footprint of the sensor are dramatically increased and such an approach is infeasible for many field applications (e.g. NQR detection of explosives \cite{garroway2001remote, cooper2016atomic}, or when the required detection distance is greater than a few centimeters\cite{array}).  As a result, the practicality of shielded AMs is limited. This motivates the development of high-performance unshielded AMs. 

The high-sensitivity operation demonstrated inside magnetic shields cannot be \emph{directly} extended to unshielded environments \cite{groeger2006}. Previous unshielded devices have only demonstrated sensitivities on (or below) the order of \SI{100}{\femto\tesla\per\sqrt\hertz} with a multi-sensor gradiometer approach and high-temperature operation \cite{cooper2016atomic, savukov2007detection, belfi2007cesium, bevilacqua2009all, keder2014unshielded, bevilacqua2016multichannel}. A gradiometric arrangement is inherently limited to the detection of rapidly decaying magnetic fields, requiring at least one sensor being in close proximity to the source. This constrains the maximum detection distance to a few \SI{}{\centi\meter}. 

Here, we report on an innovative implementation of a RF-AM for operation in unshielded environments. Our approach couples a single-sensor RF-AM and an active feedback loop stabilizing the ambient magnetic field. We show that the dominant noise contribution -- the power line noise -- is reduced by an order of magnitude and, as a result, we maintain consistent stable operation and measurement over many days \cite{deans2018machine}. We achieve sub-picotesla sensitivity near room-temperature, with a maximum sensitivity of \SI{130}{\femto\tesla\per\sqrt\hertz}. 

In contrast to gradiometric approaches, our implementation also allows long range magnetic field measurements. Retaining the ability to detect distant sources has multiple applications in surveillance, geophysics, and telecommunications \cite{gerginov2017prospects}.

We also demonstrate the wide tunability of the magnetometer operation frequency, with sub-picotesla sensitivity extending from the \SI{}{\kilo\hertz} to \SI{}{\mega\hertz} bands. Finally, we note that our device can be easily commuted between $^{85}$Rb (for higher signal-to-noise ratio, SNR) and $^{87}$Rb (for larger dynamic range). 

Our sensor's characteristics, coupled with the inherently modular nature of the setup, make it suitable for multiple applications, from portable healthcare devices to security screening, non-destructive evaluation, and industrial monitoring.

\begin{figure*}
    \includegraphics[scale=0.73]{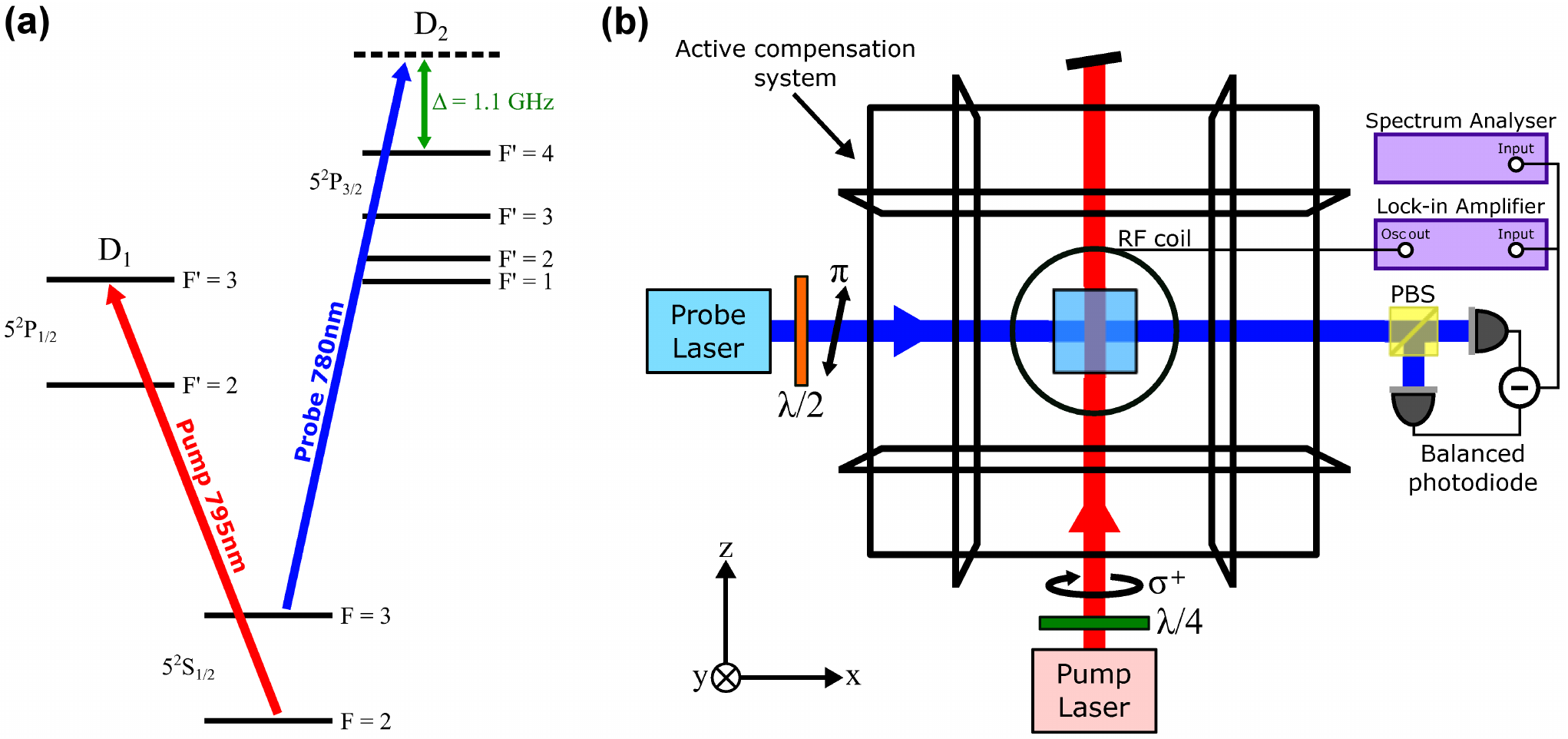}
    \hspace{0.7cm}
    \caption{\textbf{(a)} $D_1$ and $D_2$ line level diagram of $^{85}$Rb, arrows show the pump and probe excitation scheme. \textbf{(b)} Schematic of the unshielded magnetometer setup (not to scale). Rb vapor is optically pumped on the $D_1$ line with circularly polarized light. A linearly polarized probe beam crosses perpendicularly and is read-out by a polarimeter. Active compensation system (based on a PID feedback loop) maintains the desired DC magnetic field, reduces noise, and corrects field gradients -- see main text. RF coil provides a uniform calibration field, $B_{\text{RF}}$. $\lambda/2$: half-waveplate, $\lambda/4$: quarter-waveplate, PBS: polarizing beam-splitter.}
    \label{fig:Figure1}
\end{figure*} 

\section{\label{sec:Setup}Setup}

The experimental setup is shown in Fig.~\ref{fig:Figure1}. A \SI{25}{\milli\metre} cubic vapor cell of natural isotopic rubidium and \SI{20}{Torr} of N$_2$ buffer gas is temperature controlled by a \SI{1.5}{\watt} AC-current heater. This consists of thin copper wire wrapped in an anti-inductive arrangement around the cell. The heater's current is modulated by a computer-controlled H-bridge. This allows the heater frequency to be adjusted to a value far detuned from the magnetometer resonance when exploring the large dynamic range of the device, and to momentarily automatically disable the heater current during data acquisition for further noise reduction.
 
The atomic vapor is spin polarized by a circularly polarized pump laser beam locked to the $\ket{F=2} \rightarrow \ket{F'=3}$ transition of the $^{85}$Rb $D_1$ line. 

The operation frequency is set by a DC magnetic field collinear to the pump beam. This field is actively maintained by a compensation coil system. This consists of a \SI{1.2}{\metre} square 3-axes coil cage, a fluxgate magnetometer (Bartington MAG690) -- situated near the cell, and PID controllers (Stanford Research Systems SIM960). Along the pump beam axis ($\mathbf{\hat{z}}$), the bias field is provided by either a \SI{1.2}{\metre} square Helmholtz coil or a \SI{30}{\centi\metre} diameter circular Helmholtz coil (for high-field operation, not shown in Fig.~\ref{fig:Figure1}(b)). The PID acts to minimize the difference between the bias field ($B_z$, measured by the fluxgate) and the desired operation field. The current supplied to the coils is regulated by MOSFET whose gate voltage is driven by the PID output. In this way, ambient magnetic field variations and oscillating magnetic noise are actively compensated. As a result, the bias field is locked at the desired set-point. The performance of this system is evaluated in Section~\ref{sec:ActiveComp}. Magnetic field homogeneity along $\mathbf{\hat{z}}$ is further increased by field gradient compensation with a \SI{1.2}{\metre} square anti-Helmholtz coil. The transverse ambient magnetic fields (along $\mathbf{\hat{x}}$ and $\mathbf{\hat{y}}$) are zeroed -- by minimizing the Larmor frequency to the expected value for a given $B_z$ -- and can be optionally maintained at that value with two further feedback loops.

The magnetometer is calibrated with a known AC magnetic field ($B_{\text{RF}}$) provided by a pair of \SI{18}{\centi\metre} diameter Helmholtz coils in the y-direction. This field excites spin coherences between nearest-neighbor ground state Zeeman sub-levels producing a transverse atomic polarization rotation. 

The atomic precession is read out via the Faraday rotation of the plane of polarization of a linearly polarized probe beam. This beam is produced by a second laser, blue-detuned by \SI{1.1}{\giga\hertz} from the $\ket{F=3} \rightarrow \ket{F'=4}$ transition of the $^{85}$Rb $D_2$ line and crossing perpendicular to the pump beam at the centre of the cell. The overlapping region of the beams defines the sensor volume (pump beam waist \SI{5}{\milli\meter}, probe beam waist \SI{4}{\milli\meter}, volume \SI{57}{\milli\metre\cubed}). Retaining high sensitivity detection whilst significantly reducing the effective volume (compared to previously reported unshielded magnetometers) represents an important advantage in many applications. In particular, this allows an increase in the spatial resolution of magnetic field measurements. A larger sensor volume would allow a further increase in sensitivity. However, a smaller volume is used to increase the spatial resolution of magnetic field measurements in EII applications.

A polarimeter, consisting of a polarizing beam splitter and balanced photodiode, detects the probe beam rotations. The output of the photodiode is interrogated by a lock-in amplifier (LIA, Ametek DSP7280) and a spectrum analyzer (SA, Anritsu MS2718B). The internal oscillator of the LIA generates $B_{\text{RF}}$, reducing the footprint of the system.

In Section~\ref{sec:FreqRange}, we present a comparison between the high-frequency operation using $^{85}$Rb and $^{87}$Rb. In the latter case, the corresponding transitions are; pump $\ket{F=1} \rightarrow \ket{F'=2}$ transition of the $D_1$ line, probe $\ket{F=2} \rightarrow \ket{F'=3}$ transition of the $D_2$ line. As with the $^{85}$Rb, the probe beam is blue-detuned by \SI{1.1}{\giga\hertz} from the reference transition.

\section{\label{sec:Results}Results and Discussion}

\subsection{\label{sec:Optimisation}Characterization and Optimization}

A known RF calibration field ($B_{\text{RF}}$) of amplitude \SI{6}{\nano\tesla} is applied to characterize and optimize the performance of the magnetometer. The operation frequency for the optimization is chosen to be close to \SI{100}{\kilo\hertz} (bias field, $B_z=\SI{2.13e-5}{\tesla}$).

\begin{figure}[htbp]
\centering
  \includegraphics[width=0.8\linewidth]{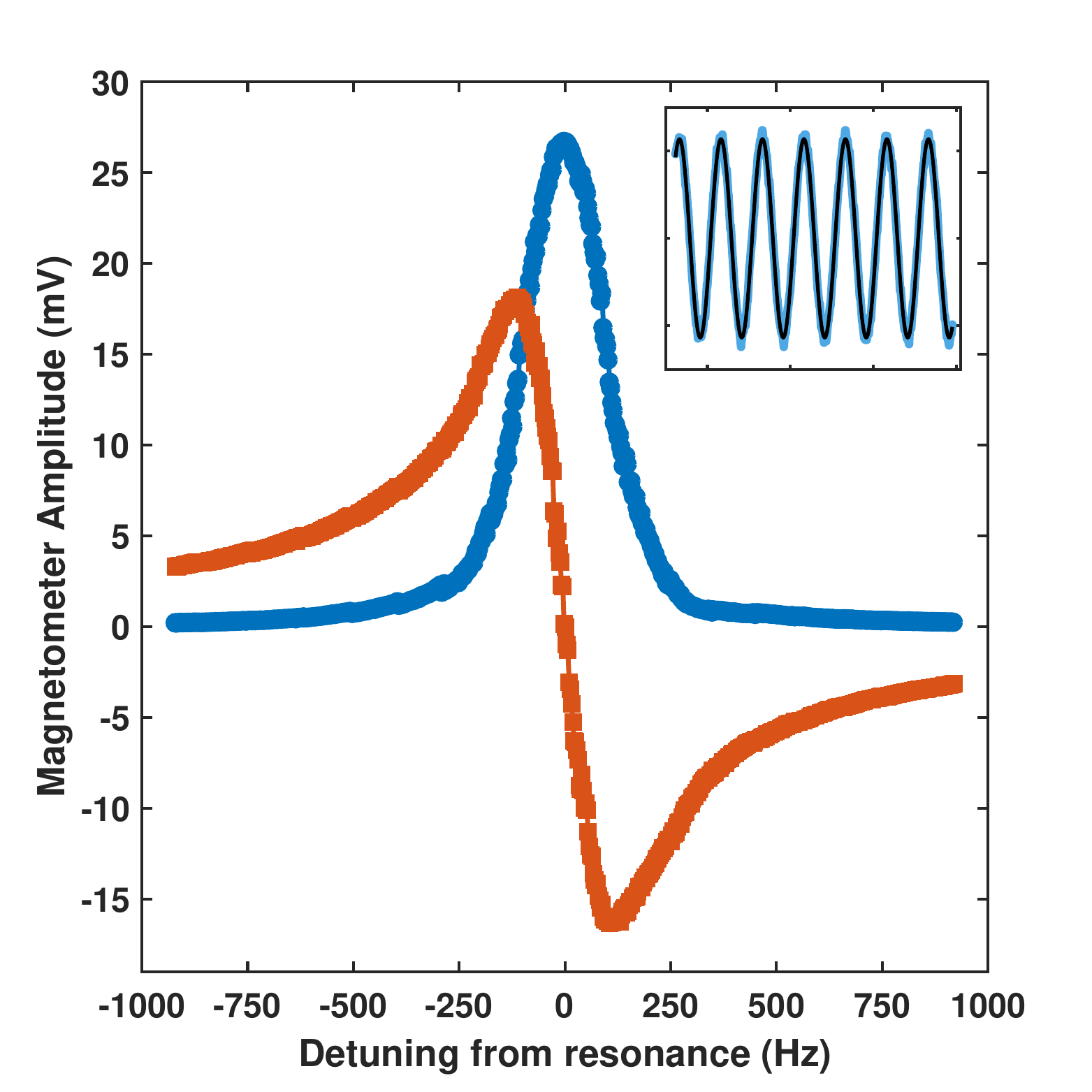}
  \caption{Typical in-phase (blue circles) and out-of-phase (red squares) response of the magnetometer near \SI{100}{\kilo\hertz} ($B_z=\SI{2.13e-5}{\tesla}$). Inset: polarimeter output at the resonant frequency fitted with sine wave of the same frequency. Near-optimized conditions: pump power \SI{400}{\micro\watt}, probe power \SI{70}{\micro\watt}, vapor cell temperature \SI{45}{\degree C}.}
  \label{fig:Figure2}
\end{figure}

To optimize the magnetometer's performance in the parameter space, we record the in-phase signal amplitude, bandwidth, and the out-of-phase gradient from the LIA response. An example resonance response showing polarization rotation as a function of RF frequency is presented in Fig.~\ref{fig:Figure2}. The typical operation bandwidth (BW, $\Gamma$) of the sensor is $\Gamma = \SI{200}{}$~--~\SI{350}{\hertz}. A BW of this scale is suitable for multiple applications, such as NQR. 

\begin{figure}[htbp]
\centering
  \includegraphics[width=0.9\linewidth]{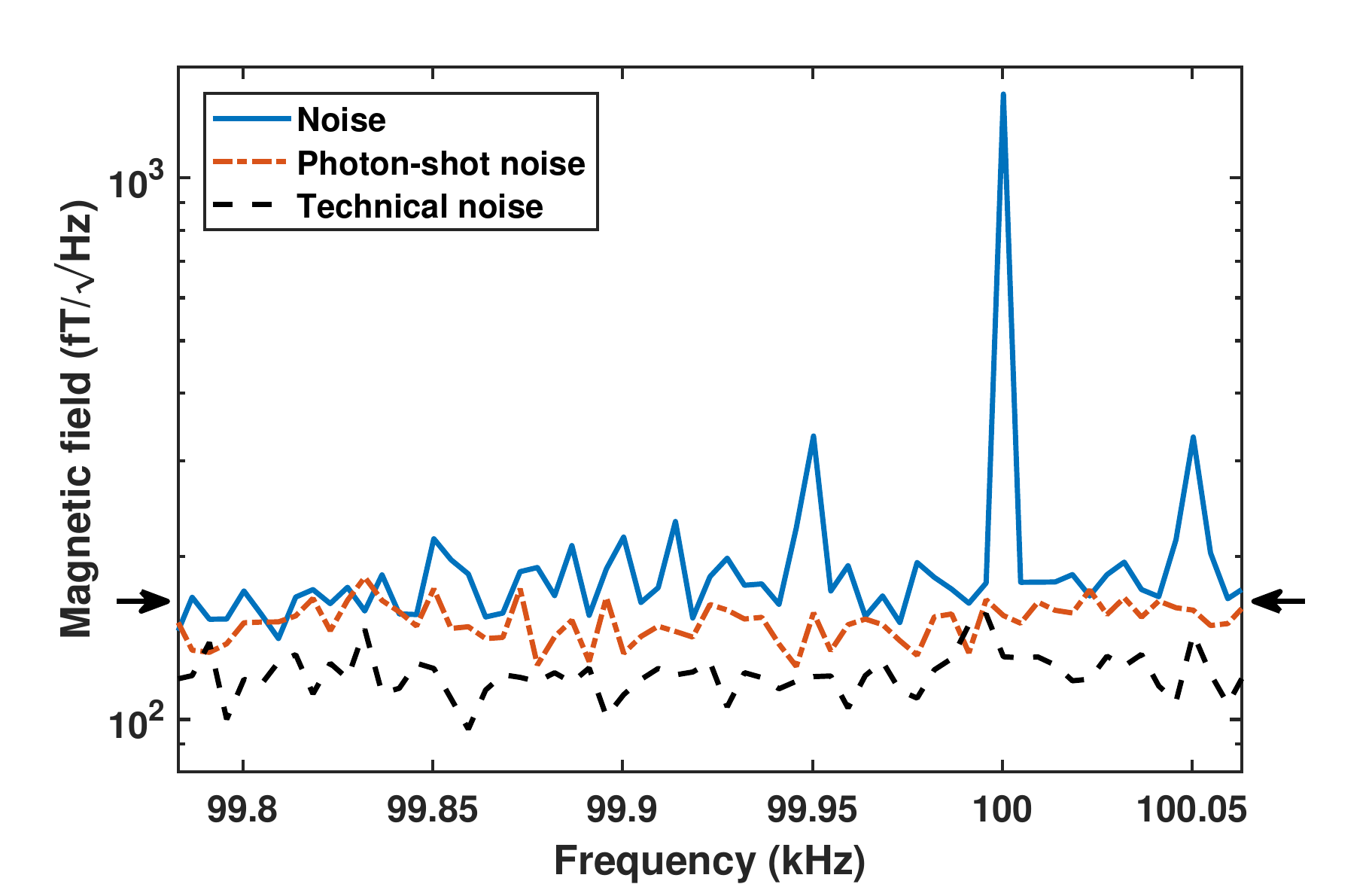}
  \caption{Magnetometer noise sources around \SI{100}{\kilo\hertz} recorded by the SA at \SI{45}{\degree C} with other parameters optimized. Total noise (solid blue line) measured with the RF field off and pump beam on. Total noise level of \SI{165}{\femto\tesla\per\sqrt\hertz} marked by arrows. Dominated by the detection of a \SI{100}{\kilo\hertz} signal and corresponding \SI{50}{\hertz} side-bands. Off resonant noise (dot-dashed red line) measured with the bias field off. Dominated by the photon-shot noise. Technical noise floor (dashed black line) measured without the probe beam.}
  \label{fig:Figure3}
\end{figure}

\begin{figure*}[ht]
  \includegraphics[scale=0.42]{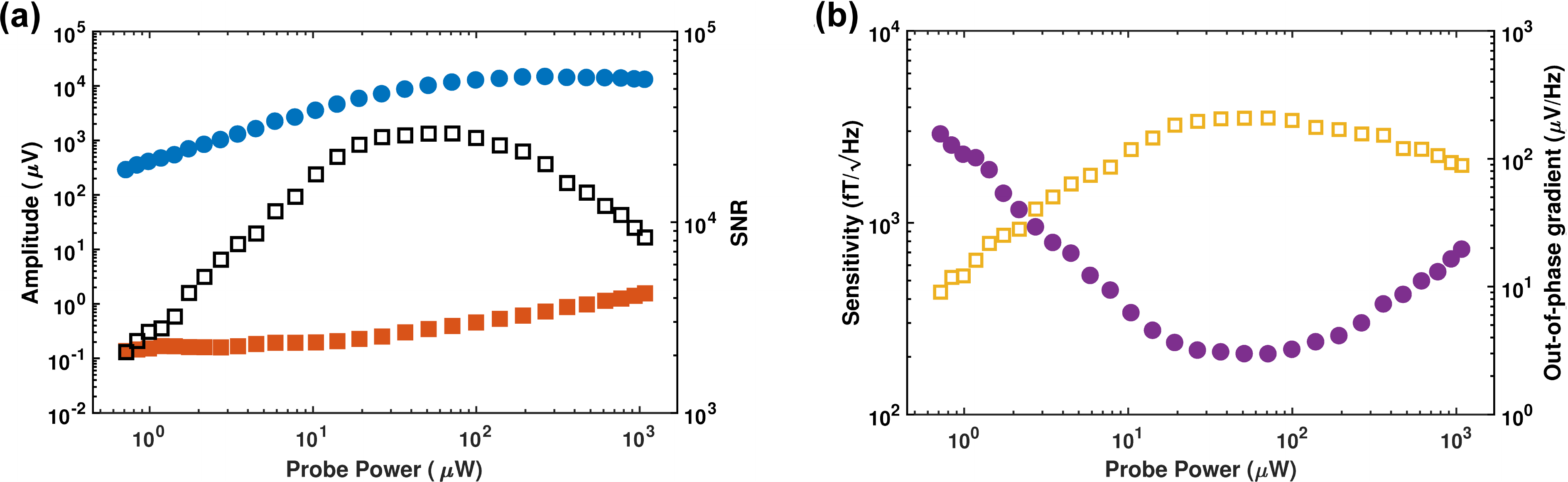}
  \hspace{0.7cm}
  \caption{\textbf{(a)} Magnetometer signal (blue filled circles), total noise -- measured with the RF field off (red filled squares), and SNR -- the ratio between the previous results (black open squares) as a function of probe beam power. Measured by the SA. \textbf{(b)} Magnetometer sensitivity (after Eq.~\ref{eqn:dbrf}, purple filled circles) and out-of-phase gradient (yellow open squares) as a function of probe beam power. All results recorded at \SI{45}{\degree C} with pump beam power \SI{400}{\micro\watt}.}
  \label{fig:Figure4}
\end{figure*}

The magnetometer performance is further evaluated by examining the signal amplitude and SNR, along with the contributing factors limiting the noise from the SA. Figure~\ref{fig:Figure3} shows the various noise components around \SI{100}{\kilo\hertz} with the magnetometer operating at \SI{45}{\degree C}. The technical noise (measured without the probe beam) arises from electrical noise in the measurement scheme. It is the dominant contribution at low probe beam powers (see Fig.~\ref{fig:Figure4}(a)). The photon-shot noise is added to this to give the off-resonant (or baseline) noise. This is recorded with the bias field detuned from the RF resonance. The photon-shot noise becomes dominant with higher probe-beam power, scaling as the square root of the power (again, see Fig.~\ref{fig:Figure4}(a)). The remaining noise terms arise from resonant noise sources. These include spin-projection noise and light-shift noise \cite{savukov2005tunable}. The total noise is recorded with the RF driving off and the bias field on (solid blue line in Fig.~\ref{fig:Figure3}). 

The total noise level describes the magnetometer sensitivity, $\delta B_{\text{1}}$. We recorded a total noise floor of \SI{165}{\femto\tesla\per\sqrt\hertz} at \SI{45}{\degree C} -- marked by the arrows in Fig.~\ref{fig:Figure3}. This level is calculated following the standard approach of calibrating the SA's vertical scale via the calibration field and the SNR\cite{savukov2005tunable}. The sensitivity is therefore given by

\begin{equation}
\delta B_{\text{1}} = \dfrac{B_{\text{RF}}}{\text{SNR}}~. \label{eqn:dbrf}
\end{equation}

Improved values would be quoted if the baseline noise was used (\SI{147}{\femto\tesla\per\sqrt\hertz}, in this case), however this is somewhat less relevant to practical applications. Therefore, we chose the total noise as the limiting factor for the AM sensitivity.

The total noise measurement reveals the detection of environmental interference at \SI{100}{\kilo\hertz} and of amplitude \SI{1.4}{\pico\tesla}. This signal, and the corresponding \SI{300}{\femto\tesla} \SI{50}{\hertz} side-bands, originate from a neighboring laboratory.

By examining the LIA and SA measurements throughout the parameter space the optimum parameter values can be set to satisfy the desired requirements of the sensor's operation -- e.g. narrowest BW, highest sensitivity. As an example, the effect of the probe beam power is shown in Fig.~\ref{fig:Figure4}. Fig.~\ref{fig:Figure4}(a) presents the results from the SA analysis of the AM output. The magnetometer amplitude (blue circles) increases with probe beam power to at maximum around \SI{300}{\micro\watt} -- decreasing slightly beyond this. The decrease at higher powers is due to the probe beam disrupting the population alignment created by the pump beam (pump beam power \SI{400}{\micro\watt}). 

The total magnetometer noise (red squares in Fig.~\ref{fig:Figure4}(a)) is recorded with the bias field on and the RF driving off, as shown in Fig.~\ref{fig:Figure3}. The SNR is computed directly from the above measurements. Figure~\ref{fig:Figure4}(b) displays the magnetometer sensitivity ($\delta B_{\text{1}}$, see Eq.~\ref{eqn:dbrf}) and the out-of-phase gradient at the resonant frequency as a function of the probe beam power. The behavior of these results is inverted as the sensitivity is inherently inversely related to the out-of-phase gradient. This follows from an alternative figure of merit for the sensitivity, $\delta B_{\text{2}} = \frac{\hbar}{g \mu_B} \frac{\Gamma}{\text{SNR}}$ -- where $\frac{\text{SNR}}{\Gamma}$ is effectively equivalent to the out-of-phase gradient \cite{pustelny2008magnetometry, lucivero2014shot}. We note that the same \SI{70}{\micro\watt} probe beam power maximizes the gradient, sensitivity, and SNR.

The optimized values at \SI{100}{\kilo\hertz} require a total optical power of only \SI{470}{\micro\watt} supplied to the sensor and \SI{29.5}{\watt} supplied to the compensation coils.

Another significant parameter to explore is the atomic vapor density. The AC current heater allowed control of the vapor cell temperature from \SI{21}{}~--~\SI{60}{\degree C}: this coresponds to densities in the range \SI{6.1e9}{}~--~\SI{2.4e11}{\per\centi\meter\cubed}. This range is significantly lower than that explored in many previous works \cite{cooper2016atomic, savukov2007detection, keder2014unshielded}. The absence of an oven reduces the size and complexity and increases the ease-of-operation in practical applications.

\begin{figure}[htbp]
\centering
  \includegraphics[width=0.9\linewidth]{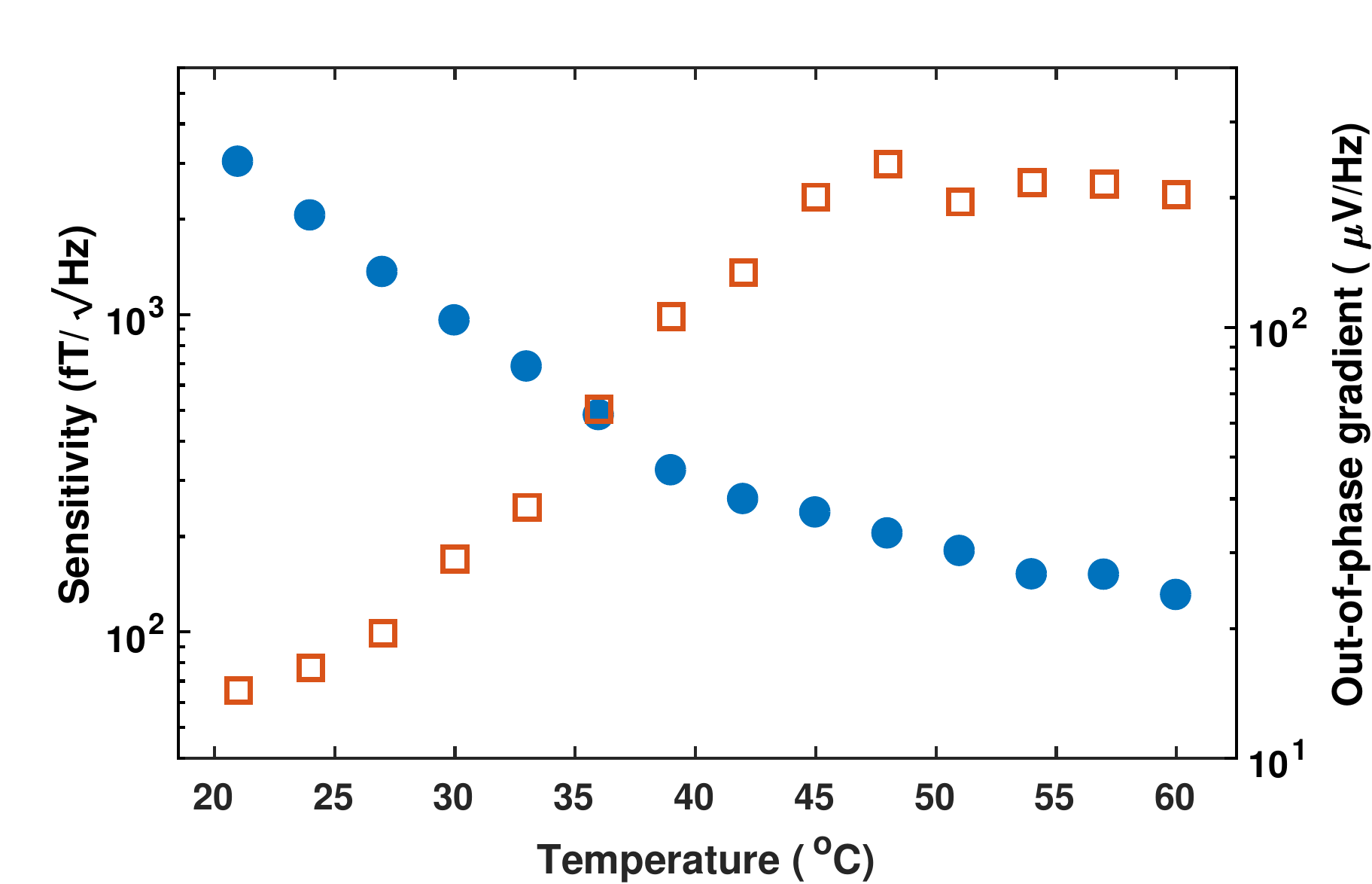}
  \caption{Magnetometer sensitivity (blue filled circles) and out-of-phase gradient (red open squares) as a function of vapor cell temperature. All results recorded at with pump beam power \SI{400}{\micro\watt} and with probe beam power \SI{70}{\micro\watt}. A peak sensitivity of \SI{130}{\femto\tesla\per\sqrt\hertz} was recorded at \SI{60}{\degree C}.}
  \label{fig:Figure5}
\end{figure}

In Fig.~\ref{fig:Figure5}, we present the magnetometer sensitivity and out-of-phase gradient as a function of temperature. Increasing the vapor density increases the sensitivity but also increases the rate of spin-exchange collisions (broadening the BW), and increases radiation trapping of the probe beam (fixed at \SI{70}{\micro\watt} across the temperature range). Both of these effects reduce the signal. Nevertheless, in the experimental range, increasing vapor density had a positive effect on the sensitivity, $\delta B_{\text{1}} = \SI{130}{\femto\tesla\per\sqrt\hertz}$ at \SI{60}{\degree C}. However, we note that the rate of improvement slowed above \SI{42}{\degree C}. This is mirrored in out-of-phase gradient where the steepest response (and also the narrowest BW) was recorded around \SI{45}{\degree C}. Sub-picotesla sensitivity was achieved with only gentle heating to \SI{29}{\degree C}.

\subsection{\label{sec:ActiveComp}Active magnetic field stabilization}

The consistent high-sensitivity operation of the magnetometer (obtained without averaging) in unshielded environments is made possible by the active compensation of stray static and oscillating magnetic fields. Operating without averaging speeds up practical applications.

The active field stabilization is based around a PID feedback loop from a fluxgate sensor. The output of the PID is used to control the current flowing in the bias field coils by controlling the gate voltage of a MOSFET. For low-field operation ($\nu < \SI{200}{\kilo\hertz}, B_{z} < \SI{4.26e-5}{\tesla}$), the current in the bias field coils can be supplied directly from the PID output. We did not record a difference in the sensor's characteristics between the MOSFET driven and PID only configurations. Furthermore, the results presented throughout this work used a single feedback loop maintaining $B_z$. The system can be easily extended to included two more loops maintaining $B_x = B_y = 0$. We found that this approach does not further improve the sensitivity though it is inherently more suited to long term data acquisition and field applications \cite{deans2018machine}.

\begin{figure}[htbp]
\centering
  \includegraphics[width=0.9\linewidth]{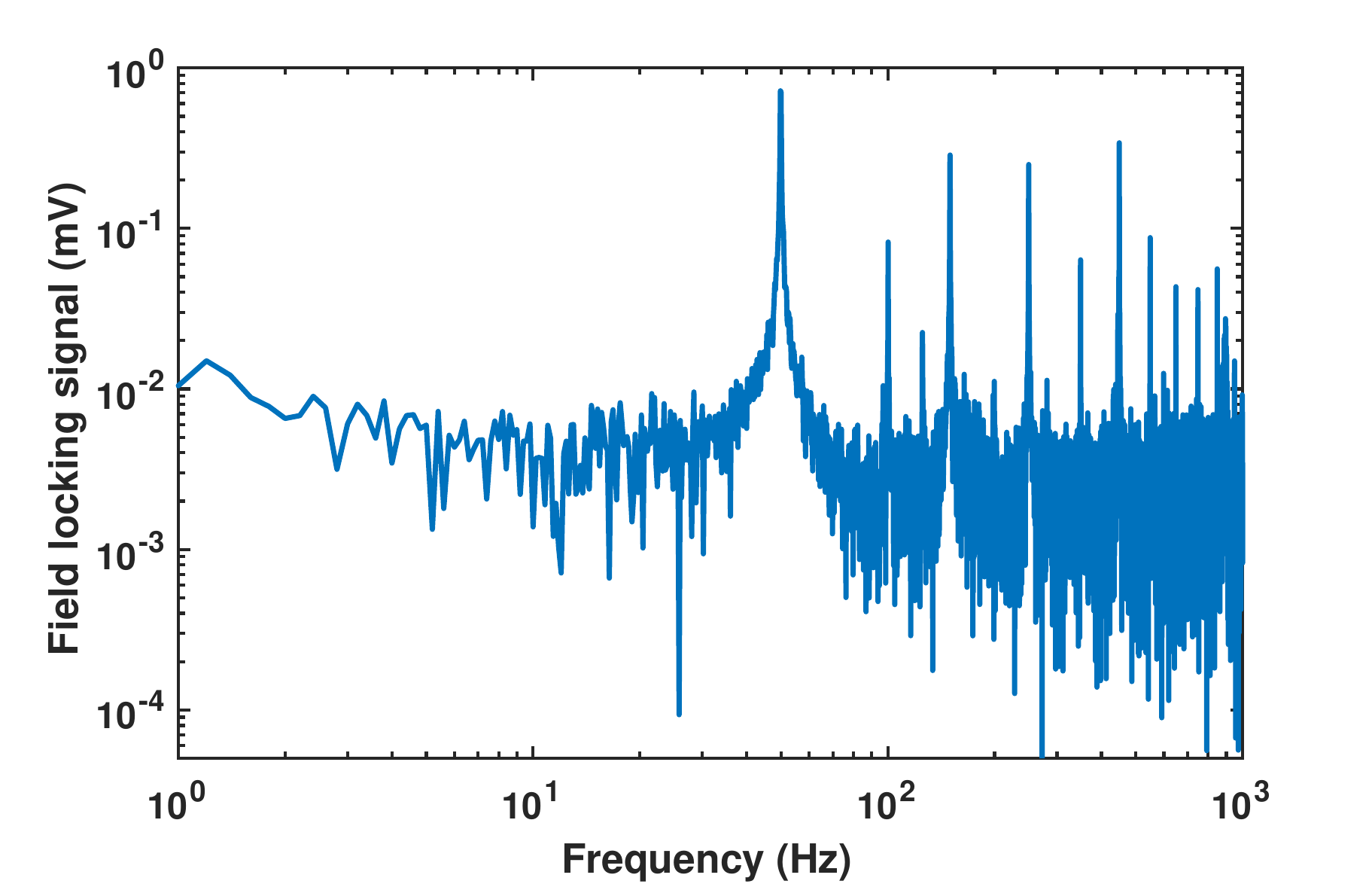}
  \caption{FFT of the PID output used to actively stabilize the bias field ($B_z$). The correction signal is dominated by \SI{50}{\hertz} noise and corresponding harmonics.}
  \label{fig:Figure6}
\end{figure}

The Fourier transform of the PID output is presented in Fig.~\ref{fig:Figure6}. This shows the frequencies that the system is working to compensate -- up to the \SI{3}{dB} bandwidth of the fluxgate at \SI{1}{\kilo\hertz}. As expected, the signal is dominated by the \SI{50}{\hertz} correction from the power line noise and its higher order harmonics.

\begin{figure}[htbp]
\centering
  \includegraphics[width=0.9\linewidth]{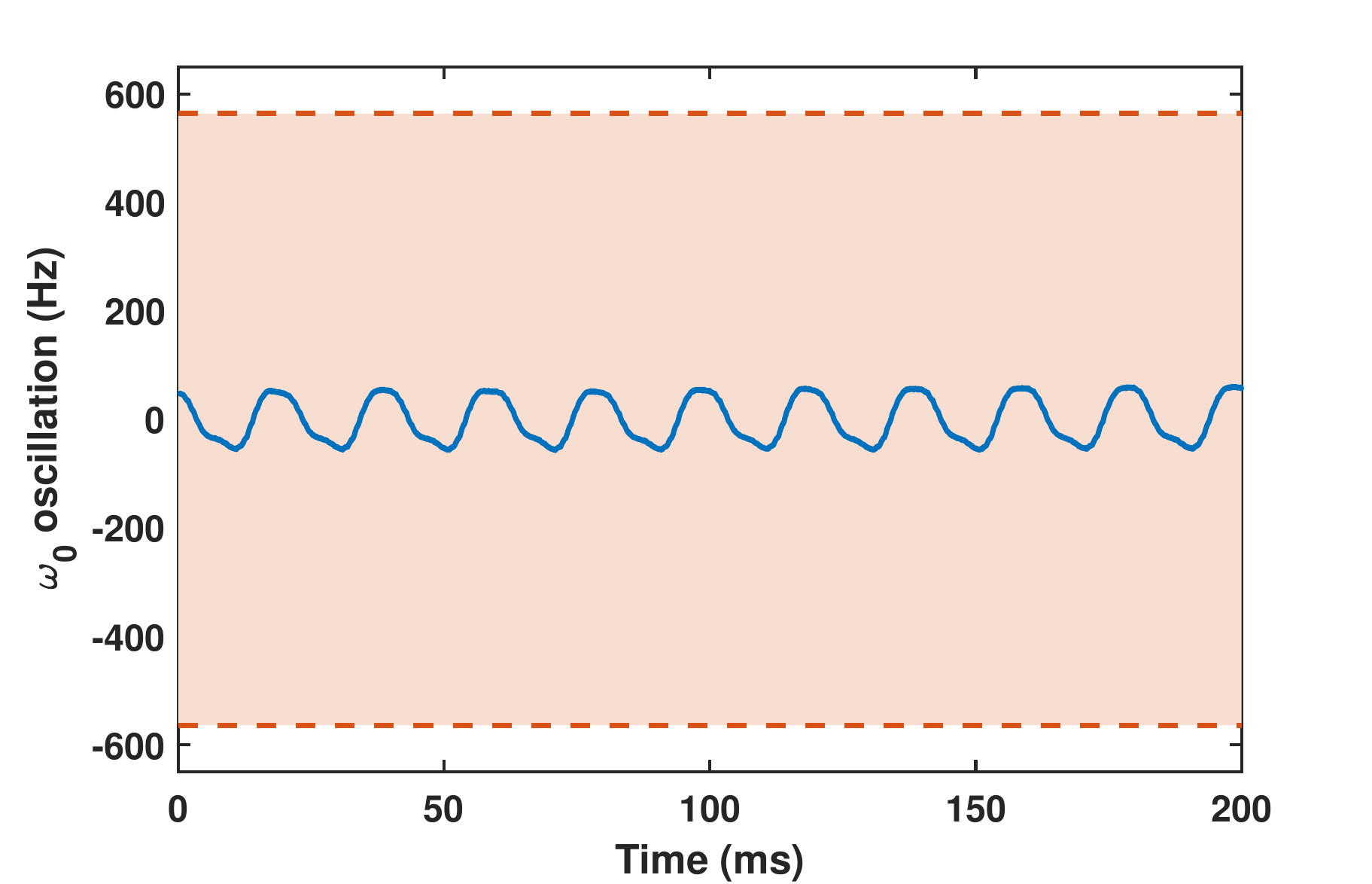}
  \caption{Oscillations in resonant frequency of the magnetometer. Active magnetic field locking reduces the \SI{50}{\hertz} noise by an order of magnitude. With field stabilization, the modulation is $\pm\SI{56}{\hertz}$ (blue line), this corresponds to $B^{\SI{50}{\hertz}}_z= \SI{11.9}{\nano\tesla}$. Without stabilization, $B^{\SI{50}{\hertz}}_z= \SI{120}{\nano\tesla}$, giving a modulation of \SI{1.1}{\kilo\hertz} -- level indicated by the shaded area.}
  \label{fig:Figure7}
\end{figure}

The \SI{50}{\hertz} noise contribution is the dominant source of noise for unshielded operation of atomic magnetometers. In this work, the sensitivity is degraded by an approximate factor 2 without the active compensation system. Consistent sensor operation is also impossible, with the resonance lineshape severely distorted and non-repeatable (see Supplementary Material~\cite{supp_material}). This is due to the distortion of $B_z$ at \SI{50}{\hertz}. We call the amplitude of this modulation $B^{\SI{50}{\hertz}}_z$. Without active compensation, this value was independently measured by two sensors and found to be \SI{120}{\nano\tesla}. For $^{85}$Rb this corresponds to an oscillation of the resonant frequency of \SI{1.1}{\kilo\hertz} (similar to those reported for other unshielded atomic magnetometers \cite{cooper2016atomic}). When the compensation system is active, we can extract the residual oscillations in the bias field directly from the LIA. This is shown in Fig.~\ref{fig:Figure7}. In these conditions the oscillations are $\pm \SI{56}{\hertz}$, this equates to an amplitude $B^{\SI{50}{\hertz}}_z = \SI{11.9}{\nano\tesla}$. The active compensation system reduces the dominant \SI{50}{\hertz} noise component by a factor ten.

During sensor operation, the effect of the noise is further reduced by time-gated data acquisition triggered directly from the power line. In this way, measurements are always taken at the same point of the \SI{50}{\hertz} oscillations - minimizing the standard deviation of measured values.

\subsection{\label{sec:FreqRange}Range of frequency tunability}

In this section we explore the range of operation frequency of the magnetometer. A broad tunability is a crucial feature for many practical applications where a range of detection frequencies are required: for example, the detection of NQR signals and tuning the penetration depth in EII.

To maintain relevance to these applications we fix all parameters at the values optimized at \SI{100}{\kilo\hertz} ($B_{\text{RF}} =\SI{6}{\nano\tesla}$, pump power \SI{400}{\micro\watt}, probe power \SI{70}{\micro\watt}, temperature \SI{45}{\degree C}). We note that tuning $B_{\text{RF}}$ across a wide frequency band requires the calibration of the RF coil across the range, although this can be performed \textit{a priori}.

We demonstrate consistent operation without averaging across three orders of magnitude: from \SI{3.5}{\kilo\hertz} to \SI{2}{\mega\hertz}, currently limited by the LIA bandwidth. This corresponds to a bias fields ($B_z$) from \SI{0.74}{\micro\tesla} to \SI{285}{\micro\tesla}. This dynamic range is significantly greater than those previously reported and confirms the suitability of our device to multiple applications.

\begin{figure}[t]
\centering
  \includegraphics[width=0.9\linewidth]{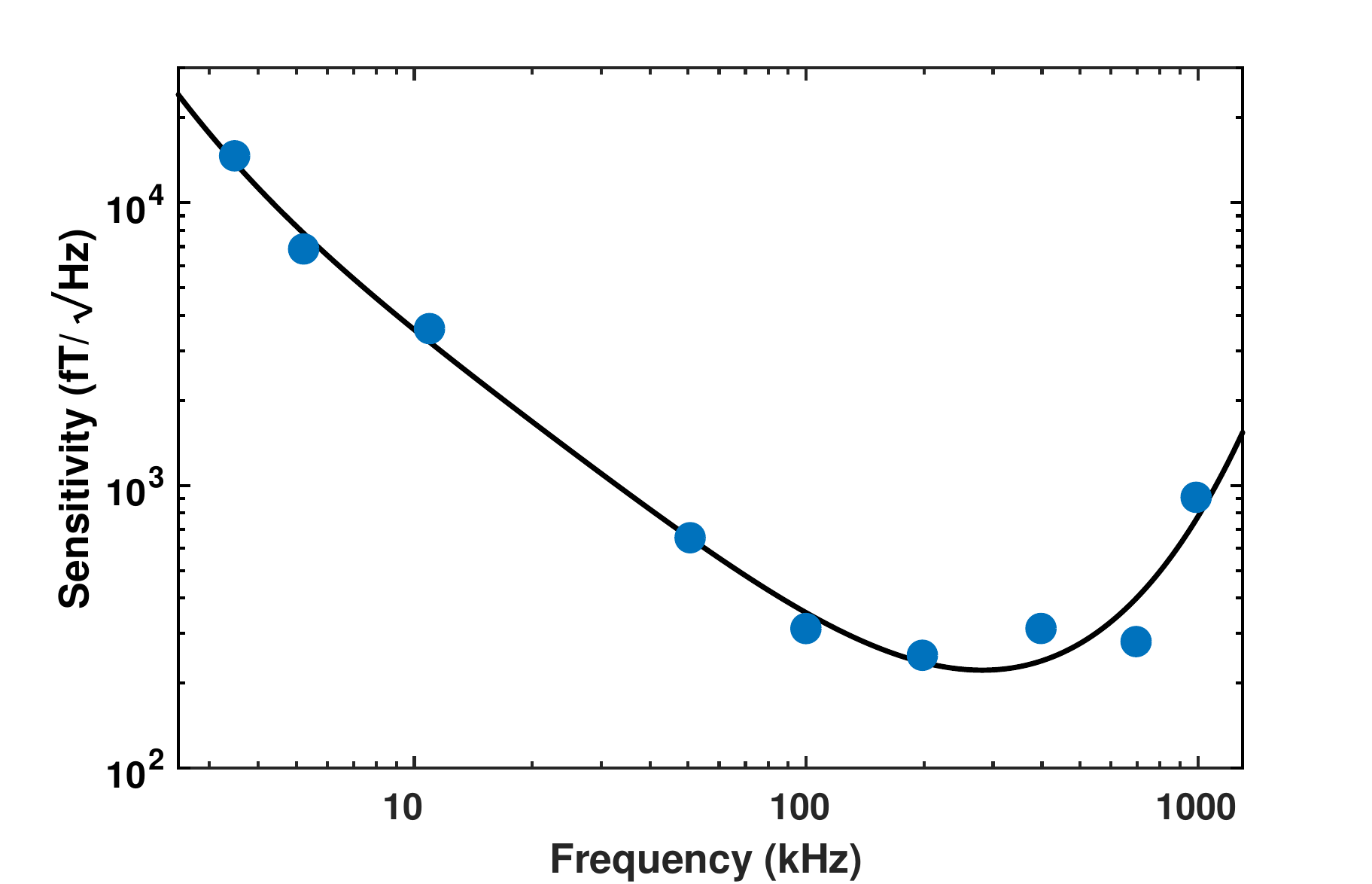}
  \caption{Magnetometer sensitivity as a function of operation frequency, demonstrating consistent operation across three orders of magnitude. Line serves to guide the eye.}
  \label{fig:Figure8}
\end{figure}

The range of frequency tunability is presented in Fig.~\ref{fig:Figure8}, in the case of the $^{85}$Rb RF-AM. We find that the magnetometer operates with sub-picotesla sensitivity from \SI{30}{\kilo\hertz} to \SI{1.1}{\mega\hertz} at \SI{45}{\degree C}. This range increases to around \SI{15}{\kilo\hertz} to \SI{1.4}{\mega\hertz} at \SI{60}{\degree C}. The maximum sensitivity is found to be in the range \SI{100}{}~--~\SI{700}{\kilo\hertz}.  At lower frequencies, an increase in the technical noise became the limiting contribution. At higher frequencies the decrease in sensitivity was due to an increasing BW. 

The largest tunability is demonstrated with $^{87}$Rb, which provides higher dynamic range, thanks to its larger gyromagnetic factor ($\gamma_{87}\approx1.49\gamma_{85}$).

\begin{figure}[htpb]
\centering
  \includegraphics[width=0.9\linewidth]{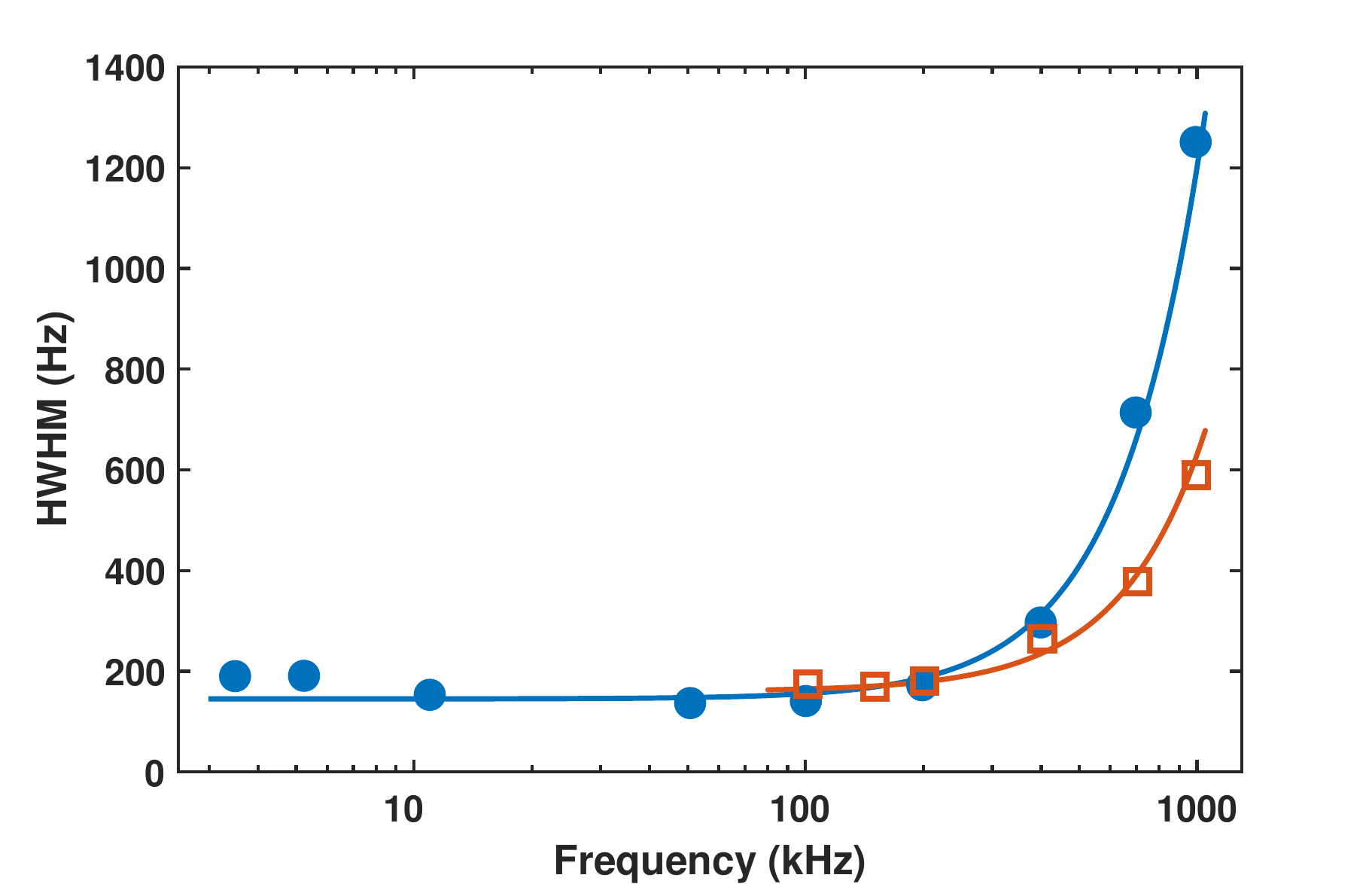}
  \caption{Magnetometer HWHM ($\Gamma/2$) as a function of frequency for both $^{85}$Rb (blue filled circles) and $^{87}$Rb (red open squares). Theoretical fits (solid lines) are calculated from the second-order Zeeman effect as a function of frequency.}
  \label{fig:Figure9}
\end{figure}

In Fig.~\ref{fig:Figure9}, the source of the broadening is explored. We compare the broadening at high $B_z$ fields for the $^{85}$Rb and the $^{87}$Rb magnetometers. The magnetometer HWHM ($\Gamma/2$) for each isotope is fitted with a function 

\begin{equation}
f(\nu) = \alpha_{85,87} + \beta_{85,87} \Phi^{85,87}(\nu),
\end{equation}

\noindent where $\Phi^{85,87}(\nu)$ is the second-order Zeeman effect as a function of operation frequency ($\nu$). $\Phi^{85,87}(\nu)$ is calculated from the Breit-Rabi formula \cite{supp_material,breit1931measurement}. The best fit parameters are: $\alpha_{85} = \SI{145}{\hertz}$, $\alpha_{87} = \SI{160}{\hertz}$, $\beta_{85}=\beta_{87}=1.6$. The strong agreement with experimental data confirms that the broadening results from second-order effects and not from the introduction of magnetic field gradients when operating in high fields. An example of the distortion of AM lineshape due to the second-order Zeeman effect is shown in the Supplementary Material~\cite{supp_material} at $\nu=\SI{2}{\mega\hertz}$, where the $^{87}$Rb AM exhibits a sensitivity of \SI{4.3}{\pico\tesla\per\sqrt\hertz}.

\section{\label{sec:Conclusions}Conclusions}

We have demonstrated a single-channel unshielded RF-AM, operating with sub-picotesla sensitivity near room temperature. Our approach is based on the active locking the magnetic field that the sensor operates in. We measured a resulting order of magnitude reduction in the power-line noise. The single-sensor approach is applicable to the detection of both local and remote magnetic field sources. The small effective volume of the sensor increases the spatial resolution of the measurement and leads to an improved imaging resolution in EII applications. We have shown that the operation frequency of the sensor is tunable across three orders of magnitude. The performance is degraded, in the \SI{}{\mega\hertz} range, by the second-order Zeeman effect. Our current range is only limited by the bandwidth of the LIA. Operating without any passive magnetic shielding, the demonstrated flexibility in operating conditions, and low optical power greatly increase the applicability of our device to practical applications.

\section*{Supplementary Material}
See \href{http://aip.scitation.org/rsi/}{supplementary material} for information on operation without active field stabilization, and the derivation of the second-order Zeeman effect fit ($\Phi^{85,87}(\nu)$) and its impact on the RF-AM response.

\begin{acknowledgments}
The authors would like to thank Prof.~Valerio Biancalana (University of Siena), Dr~Witold Chalupczak (NPL), Dr~Yordanka Dancheva (University of Siena), and Dr~Rafal Gartman (NPL) for their stimulating discussions.

The authors acknowledge support from the UK Quantum Technology Hub in Sensing and Metrology, Engineering and Physical Sciences Research Council (EPSRC) (EP/M013294/1). Cameron Deans acknowledges support from the Engineering and Physical Sciences Research Council (EPSRC) (EP/L015242/1).
\end{acknowledgments}

\bibliographystyle{aipnum4-1}

\onecolumngrid
\pagebreak
\newpage

\title[]{Supplementary Material for ``Sub-picotesla widely-tunable atomic magnetometer operating at room-temperature in unshielded environments''}

\author{Cameron Deans}
\author{Luca Marmugi}
\email{l.marmugi@ucl.ac.uk}
\author{Ferruccio Renzoni}
\affiliation{Department of Physics and Astronomy, University College London, Gower Street, London WC1E 6BT, United Kingdom}

\date{\today}

\maketitle

\onecolumngrid
Supporting material for ``Sub-picotesla widely-tunable atomic magnetometer operating at room-temperature in unshielded environments'' by C. Deans \textit{et al.} .

\section{\label{sec:Intro}Operation without active field stabilization}

\begin{figure}[htbp]
\centering
  \includegraphics[width = 0.55\linewidth]{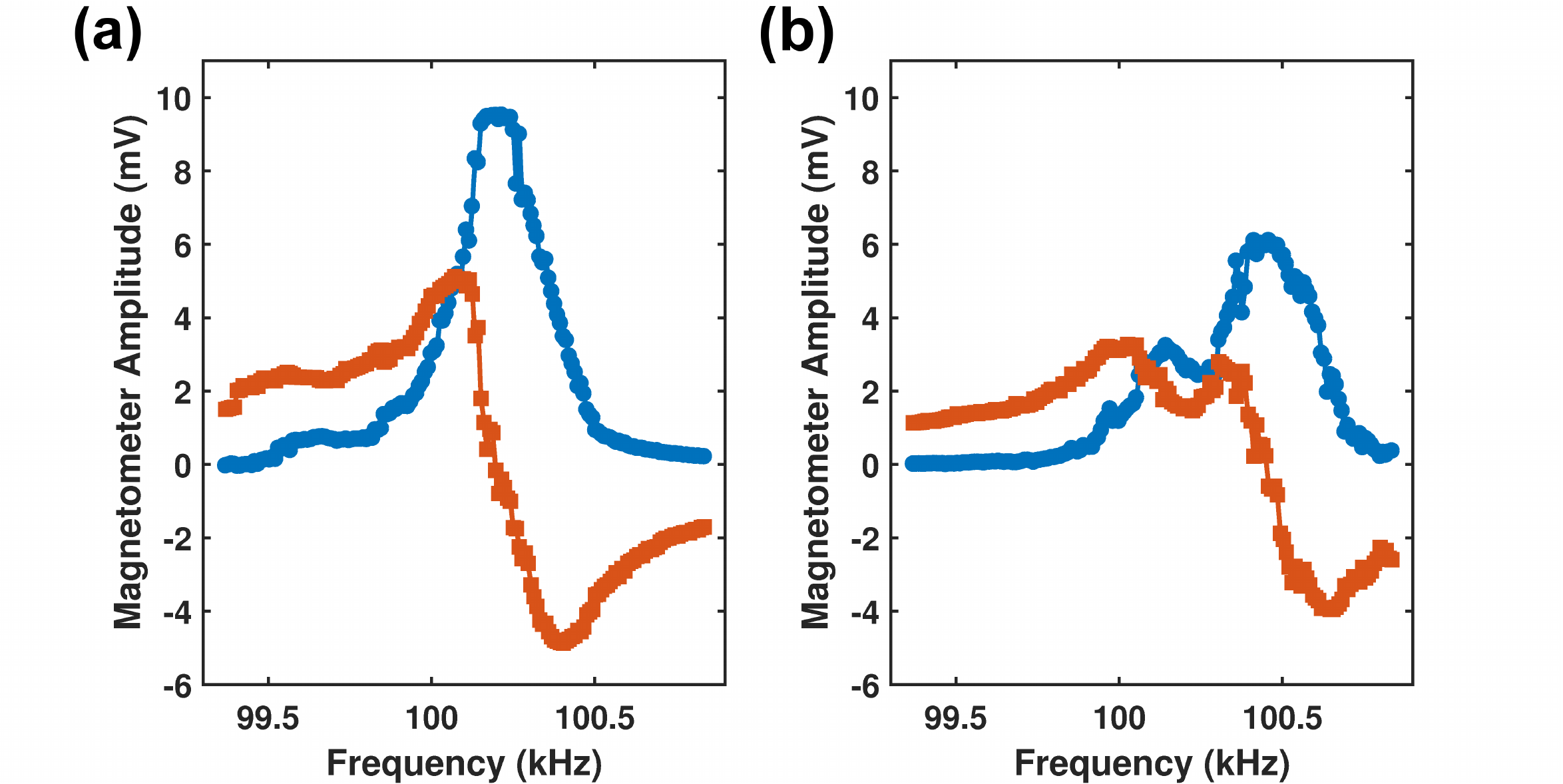}
  \caption{Typical in-phase (blue circles) and out-of-phase (red squares) responses of the magnetometer without magnetic field compensation. \textbf{(a)} and \textbf{(b)} are consecutive traces, both recorded within a \SI{30}{\second} period. The response is non-repeatable and severely distorted in comparison to the field-stabilized case (Figure~3, main text).}
  \label{fig:supp1}
\end{figure}

The sensitivity of our sensor is reduced by an approximate factor 2 without the active compensation system stabilizing the magnetic field. In addition, consistent operation would be impossible without averaging. This is because the of the distortion of the resonance lineshape, demonstrated in Figure~\ref{fig:supp1}.  

The results in Figure~\ref{fig:supp1} also show how detrimental it would be to simply remove magnetic shielding from a standard AM designed to operate in screened environment,  if no countermeasures of the type described in the main text are taken.

\section{Second-order Zeeman effect as a function of operation frequency}

The second-order Zeeman effect splits the RF resonances corresponding to specific transitions between neighboring ground state magnetic sub-levels. This becomes more important at higher bias fields (i.e. higher operation frequencies).

Transitions energies for each Zeeman state as a function of the applied magnetic field are described by the Breit-Rabi equation \cite{breit1931measurement}.

\begin{equation}
W(\ket{F,m_\text{F}}) = - \frac{\Delta W}{2(2I+1)} + g_{I}\mu_{B}m_{F}B \pm \frac{\Delta W}{2} \sqrt{1+\frac{4m_{F}}{2I+1}x + x^{2}}.
\label{eqn:briet-rabi}
\end{equation}

\noindent In the above equation: $x = \frac{(g_{J}-g_{I})\mu_{B}}{\Delta W}B$ is the dimensionless field strength parameter, the choice of sign ($\pm$) is for $F_{\pm}=I\pm \frac{1}{2}$, and $\Delta W/h$ is the $F_{\pm}$ hyperfine splitting (in \SI{}{Hz}). The constants are: the Bohr magneton $\mu_{B}$, the nuclear g-factor $g_{I}$, and the  angular momentum Land\'{e} g-factor $g_{J}$.

\begin{figure}[htbp]
\centering
\includegraphics[width = 0.6\linewidth]{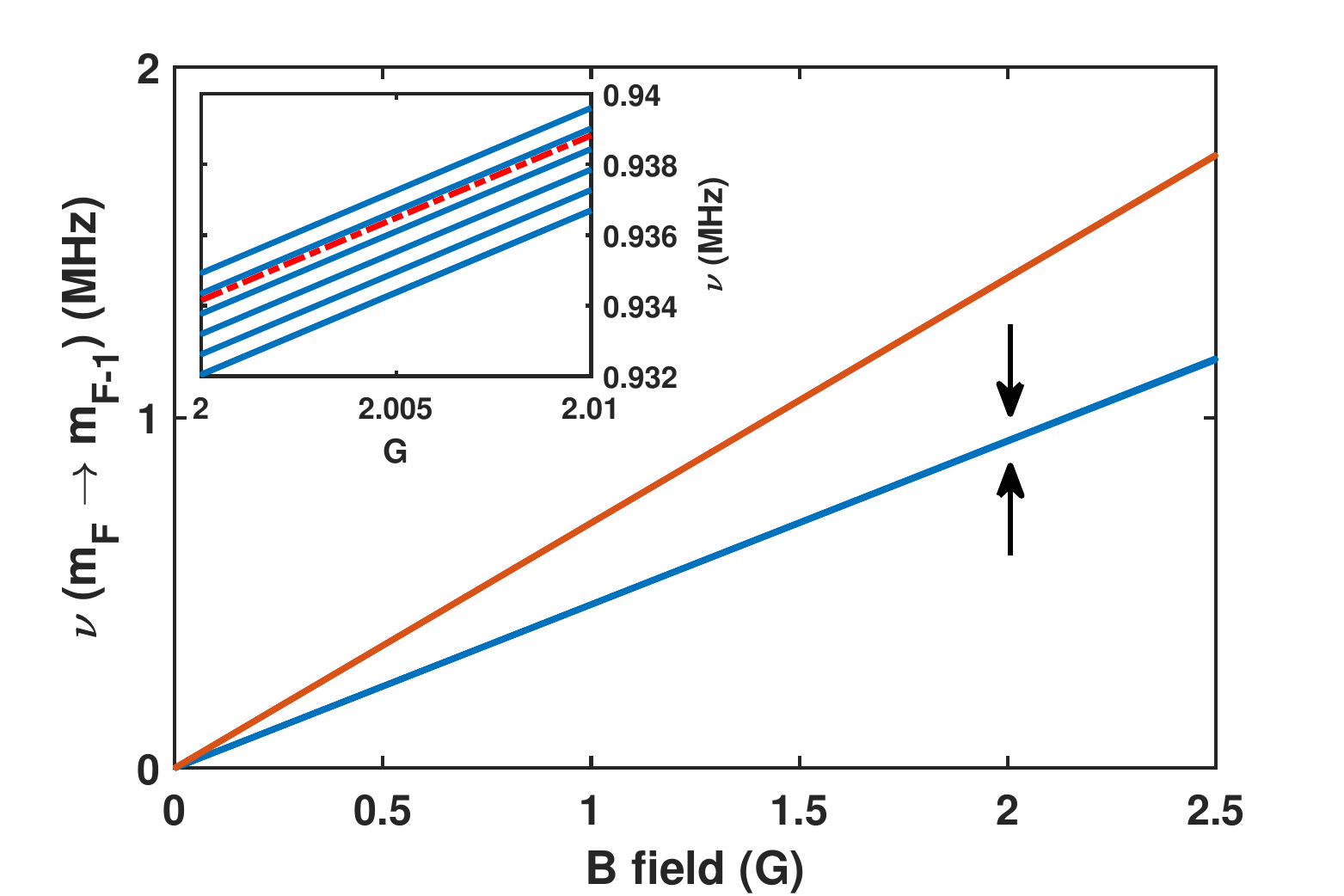}
\caption{Calculated magnetometer operation frequency for $^{85}$Rb (blue) and $^{87}$Rb (red). Inset: Magnified transitions of $^{85}$Rb split due to second-order effects (position of inset marked by arrows). Good agreement with linear approximation (dot-dashed red line).} 
\label{fig:supp2}
\end{figure}

Equation~\ref{eqn:briet-rabi} allows the calculation of the operation frequency of the RF-AM in a given magnetic field. The resulting operation frequencies for $^{85}$Rb and $^{87}$Rb (in the regime explored in the main text) are plotted in Figure~\ref{fig:supp2}. As expected, the transition frequencies are linear and in good agreement with the linear Zeeman effect - in this regime (Inset: dot-dashed red line). Nevertheless, the exact transition frequency for each pair of nearest-neighbor Zeeman sub-levels is different due to second order effects (magnified in Inset). 

\begin{figure}[htbp]
\centering
\includegraphics[width = 0.6\linewidth]{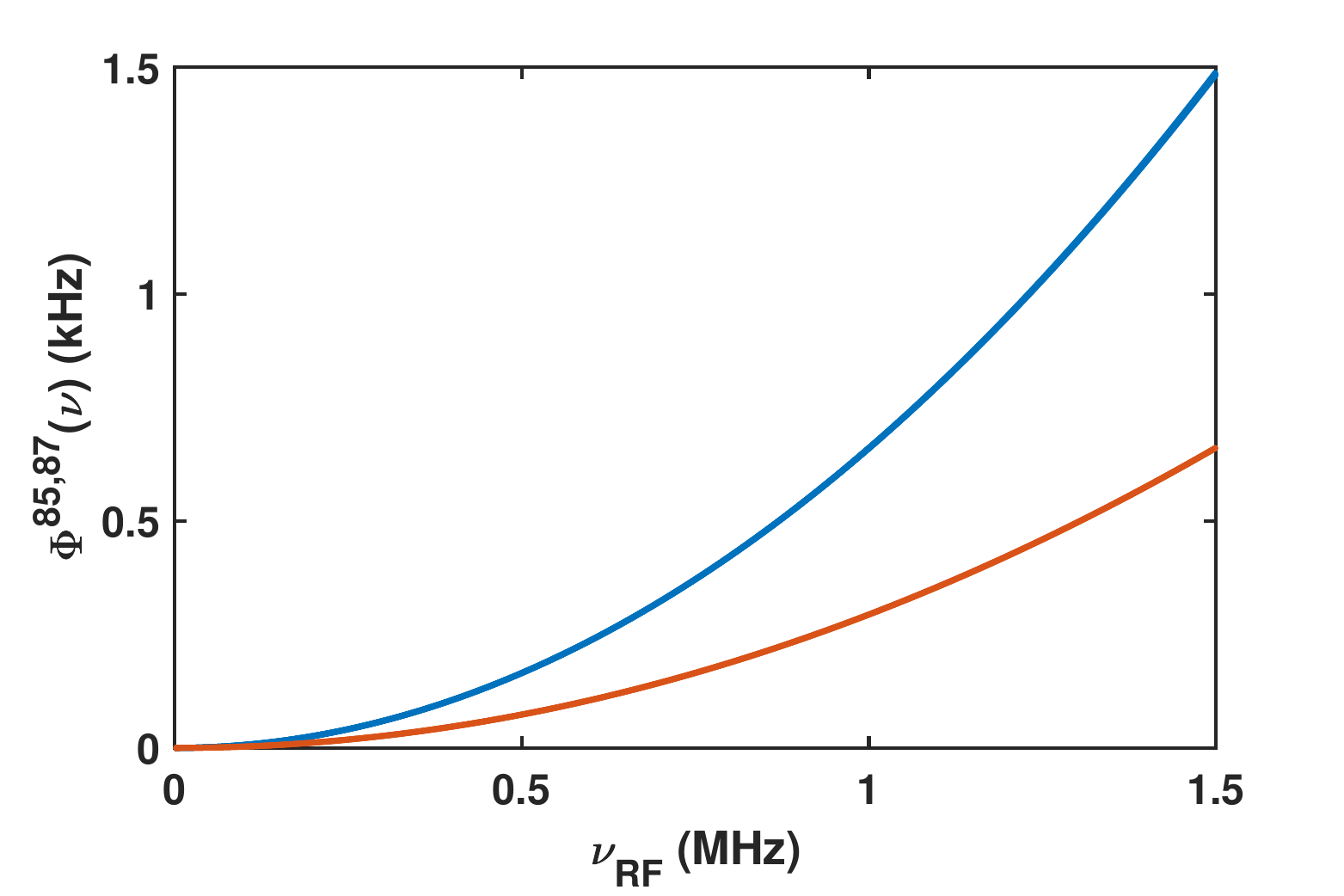}
\caption{Second-order Zeeman splitting of magnetic sub-levels as a function of operation frequency for $^{85}$Rb (blue) and $^{87}$Rb (red). These calculations are referred to as $\Phi^{85,87}(\nu)$ in the main text.} 
\label{fig:supp3}
\end{figure}

We extract the second-order contribution by subtracting neighboring lines in Figure~\ref{fig:supp2} and converting the x-axis from magnetic field to operation frequency. This gives the second-order Zeeman effect as a function of operation frequency. The resulting calculations are plotted in Figure~\ref{fig:supp3}. They are referred to as $\Phi^{85,87}(\nu)$ and used as theoretical fits in Figure~7 of the main text. 

\begin{figure}[htbp]
\centering
\includegraphics[width = 0.5\linewidth]{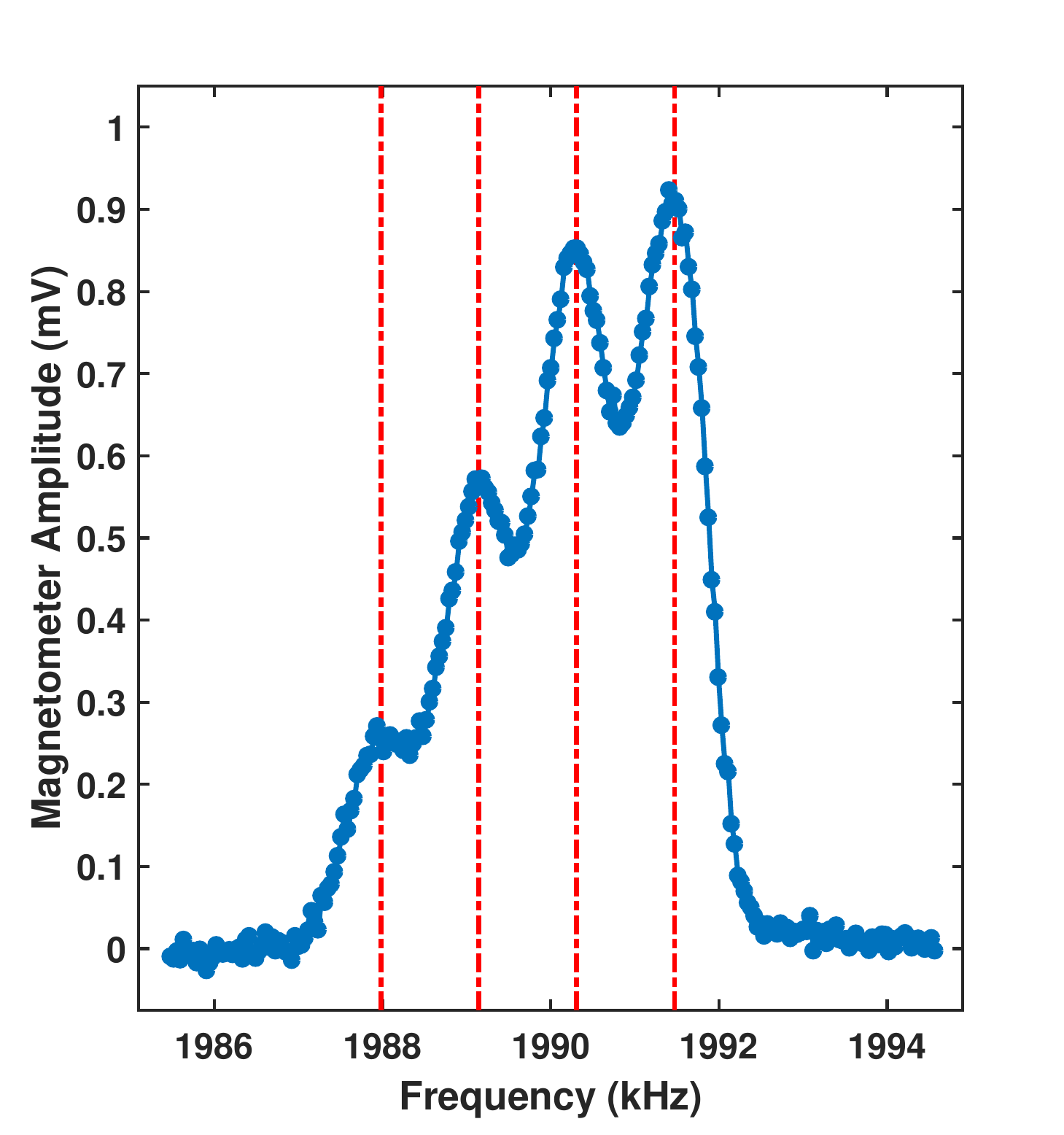}
\caption{Second-order Zeeman splitting of magnetometer resonance - $^{87}$Rb magnetometer operating around \SI{2}{\mega\hertz} (\SI{1991}{\kilo\hertz}). Marked peaks and splittings (red lines) are calculated from $\Phi^{87}(\nu)$.} 
\label{fig:supp4}
\end{figure}

The second-order broadening of the magnetometer resonance is further confirmed in Fig.~\ref{fig:supp4}, where the RF resonance is presented for the magnetometer operating on $^{87}$Rb near \SI{2}{\mega\hertz} ($B_z = \SI{285}{\micro\tesla}$). Here, the individual RF transitions between the magnetic sub-levels are clearly separated. Their separation matches that predicted from the calculation of $\Phi^{87}(\nu)$ above (dashed lines in Fig.~\ref{fig:supp4}).

\end{document}